\documentclass[12pt]{iopart} 
\usepackage{amsfonts} % needed for bold Greek, Fraktur, and blackboard bold
\usepackage{graphicx} % needed for figures
\usepackage{gensymb}
\usepackage{hyperref}

\begin{document}

% Be sure to use the \title, \author, \affiliation, and \abstract macros
% to format your title page.  Don't use lower-level macros to  manually
% adjust the fonts and centering.

\title{Significance of a one-degree Celsius increase in global temperature}
% In a long title you can use \\ to force a line break at a certain location.

\author{J S Gagnon}
\address{Department of Physics, Norwich University, Northfield VT, U.S.A.}
\ead{jgagnon6@norwich.edu}

\begin{abstract}
The Intergovernmental Panel on Climate Change reports indicate that the global mean temperature is about one-degree Celsius higher than pre-industrial levels, that this increase is anthropogenic, and that there is a causal relationship between this higher temperature and an increase in frequency and magnitude of extreme weather events.  This causal relationship seems at odds with common sense, and may be difficult to explain to non-experts.  Thus to appreciate the significance of a one-degree increase in global mean temperature, we perform back-of-the-envelope calculations relying on simple physics.  We estimate the excess thermal energy trapped in the climate system (oceans, land, atmosphere) from a one-degree Celsius increase in global mean temperature, and show that it is thousands of times larger than the estimated energy required to form and maintain a hurricane.  Our estimates show that global warming is forming a very large pool of excess energy that could in principle power heatwaves, heavy precipitation, droughts, and hurricanes.  The arguments presented here are sufficiently simple to be presented in introductory physics classes, and can serve as plausibility arguments showing that even a seemingly small increase in global mean temperature can potentially lead to extreme weather events.
%It took the work of thousands of scientists worldwide and the analysis of large data sets to reach this conclusion, making it understandably difficult to see the link between such a small temperature increase and extreme weather events.
\end{abstract}

\noindent{\it Keywords\/}: Global warming, hurricanes, back-of-the-envelope

\submitto{ejp}

\maketitle % title page is now complete

%compare this to the estimated energy required to form and maintain a hurricane.  These order-of-magnitude calculations indicate that this excess thermal energy is hundreds of times larger than the energy required to form and maintain a hurricane over its lifetime.  Since the excess energy in the atmosphere corresponds to only $1\%$ of the excess energy in the climate system (atmosphere, oceans, land), 

%%%%%%%%%%%%%%%%%%%%%%%%%%%%%%%%%%%%%%%%%%%%%%%%%%%%%%%%%%%%%%%%%%%%%%%%%%%%%%
\section{Introduction}
%%%%%%%%%%%%%%%%%%%%%%%%%%%%%%%%%%%%%%%%%%%%%%%%%%%%%%%%%%%%%%%%%%%%%%%%%%%%%%

The latest reports from the Intergovernmental Panel on Climate Change (IPCC)~\cite{IPCC_2021_PHysical_Science_Basis} indicate that the global mean surface temperature and global surface air temperature are about $1\degree$C higher than pre-industrial levels, and that this increase is related to human activities.  This increase in mean global temperature is considered a serious long term threat to humankind, and the importance of this problem has been recognized by the Nobel committee, who awarded the 2021 Nobel Prize in Physics to pioneers of climate change science~\cite{Lopatka_2001}.  Indeed, one of the goals of the $26^{\rm th}$ UN Climate Change Conference of the Parties (COP26) was to encourage countries to reduce their emissions of greenhouse gases in order to keep the increase in global mean temperature below $1.5\degree$C.

\vspace{\baselineskip}

The same reports from the IPCC indicate that there is an increase in frequency and magnitude of extreme weather events, such as heatwaves, heavy precipitation, droughts,
and tropical cyclones~\cite{IPCC_2021_PHysical_Science_Basis}.  Now there seems to be an apparent clash between such a seemingly small increase in global temperature and such extreme weather events.  It is an often heard comment (among people from the general public) that ``a $1\degree$C increase does not seem to be a lot'', thus why should we take drastic measures to keep the increase in the global mean temperature under $1.5\degree$C?

\vspace{\baselineskip}

It is no wonder that such comments can be heard in the general public.  The science of climate change and event attribution is very complex.  The system studied (i.e. oceans, land, atmosphere) has a huge number of degrees of freedom, is highly nonlinear with numerous feedback mechanisms, and highly out-of-equilibrium (due to external forcing).  In addition, to find the fingerprint of human induced changes, it is necessary to extricate the natural variability of various data sets over many orders of magnitude in spatial and temporal scales, and compare with cutting edge numerical simulations.

\vspace{\baselineskip}

A robust way of quantifying global warming on interannual-to-decadal timescales is to measure the accumulation of energy in the Earth's climate system (oceans, land, atmosphere)~\cite{IPCC_2021_PHysical_Science_Basis,vonSchuckmann_etal_2020}.  A good starting point in this endeavor is the energy balance equation~\cite{Pierrehumbert_2010,NASA_earth_energy_budget}:
\begin{eqnarray}
\label{eq:Energy_budget}
\Phi_{\rm in} - \Phi_{\rm out} & = & \Delta \Phi_{\rm imbalance},
\end{eqnarray}
where $\Phi_{\rm in}$ is the incoming energy flux from ``short'' wavelength radiation (ultraviolet, visible, near infrared) from the Sun, $\Phi_{\rm out}$ is the outgoing energy flux from ``long wavelength'' radiation (thermal infrared) from the top of Earth's atmosphere, and $\Delta \Phi_{\rm imbalance}$ is the flux imbalance between those two quantities.  When $\Phi_{\rm in} = \Phi_{\rm out}$ (i.e. $\Delta \Phi_{\rm imbalance} = 0$), the amount of energy entering and leaving are equal, and no excess energy accumulates in the Earth's climate system.  However, if $\Phi_{\rm in} > \Phi_{\rm out}$ due to some physical process, then the nonzero flux imbalance $\Delta \Phi_{\rm imbalance}$ produces an accumulation of energy in the Earth's climate system over time, which ultimately heats it up.  A large part of the complexity of climate change science lies in the measurement and computation of the quantities in Eq.~\ref{eq:Energy_budget}. 

\vspace{\baselineskip}

Based on Eq.~\ref{eq:Energy_budget}, two different strategies can be used to measure the accumulation of energy in the Earth system.  The first strategy is to directly measure the incoming radiation from the Sun ($\Phi_{\rm in}$) and outgoing radiation from the top of the atmosphere ($\Phi_{\rm out}$) using satellite imagery (see \cite{Dewitte_Clerbaux_2017} for a review), and integrate it over the surface of the Earth and over time.  Various factors make these measurements complicated.  For instance, part of the incoming solar radiation is reflected back into space, but the amount varies in space and time depending on cloud cover and the type of reflecting surface involved (e.g. forest, ice)~\cite{Coakley_2003}.  The outgoing radiation is also affected by many factors (e.g. clouds~\cite{Dewitte_Clerbaux_2018}), and in particular by the composition of the atmosphere.  The presence of certain gases (e.g. water vapor, CO$_{2}$, methane) in sufficient concentration in the air increases the opacity of the atmosphere to outgoing long wavelength radiation~\cite{Pierrehumbert_2010}, thus decreasing $\Phi_{\rm out}$ with respect to $\Phi_{\rm in}$, resulting in a radiative forcing of Earth's system.  As a result of this (net positive) radiative forcing, the planet tries to restore radiative equilibrium by increasing its global temperature, and thus its cooling via blackbody radiation (i.e. changing $\Phi_{\rm out}$ such that $\Delta \Phi_{\rm imbalance}$ becomes zero again).  The timescale over which the climate responds to this radiative forcing depends on the planet's thermal inertia and various feedback mechanisms~\cite{Hansen_etal_2005a,Hansen_etal_2005b}.  Consequently, it might take centuries for the climate to fully respond to a radiative forcing.  

\vspace{\baselineskip}

Since the Earth has been in radiative imbalance for at least 50 years, excess energy has been accumulating in the climate system.  This energy goes into warming the oceans, the land, the atmosphere, and the cryosphere (ice).  Thus a second strategy of quantifying global warming is to perform in situ and remote measurements of temperature and other quantities, in order to directly compute the total change in global energy content of the climate system (see \cite{vonSchuckmann_etal_2020} for a review).  This includes temperature measurements at various ocean depths (e.g. \cite{vonSchuckmann_etal_2005}), atmospheric temperature and air velocity measurements~(e.g. \cite{Kobayashi_etal_2015}), and borehole temperature profiles (e.g. \cite{Pickler_etal_2018}).  The change in total energy content of the climate system (obtained from direct temperature measurements) is related in a complicated manner to the flux imbalance $\Delta \Phi_{\rm imbalance}$ integrated over space and time, as they are both related to conservation of energy.  According to the IPCC reports~\cite{IPCC_2021_PHysical_Science_Basis}, both methods give the same order of magnitude increase in global energy content in the period 1971-2018 compared to pre-industrial levels ($3.40\times 10^{23}$ J for radiative forcing and $4.34 \times 10^{23}$ J for direct temperature measurements), with $91\%$ soaked up by oceans, $5\%$ by land, $1\%$ by the atmosphere, and $3\%$ used up to melt ice~\cite{vonSchuckmann_etal_2020}.

\vspace{\baselineskip}

As one can see, quantifying global warming is a major undertaking, requiring vast sets of data coming from monitoring stations all over the world as well as satellite imagery, analyzed over many decades in time.  Details about each contribution to radiative forcing are crucial in order to find the fingerprints of humans in global warming.  The goal of this paper is to present order-of-magnitude estimates of the excess energy in the climate system that are accessible to an introductory physics class.  The estimates are based on the ``1$\degree$C increase in global temperature'' popularized by the media, with the intention of making students and the general public appreciate the significance of such a small increase in global mean temperature.  Another goal of the paper is to show that this excess energy is much larger than the energy required to power an extreme weather event (in this case a hurricane), showing that it is at least plausible that such a small increase in global temperature could lead to such extreme events.  The arguments presented here are simple enough to be incorporated into an introductory physics class.

%%%%%%%%%%%%%%%%%%%%%%%%%%%%%%%%%%%%%%%%%%%%%%%%%%%%%%%%%%%%%%%%%%%%%%%%%%%%%%
\section{Global increase in the energy content of the climate system}
\label{sec:Global_increase_energy}
%%%%%%%%%%%%%%%%%%%%%%%%%%%%%%%%%%%%%%%%%%%%%%%%%%%%%%%%%%%%%%%%%%%%%%%%%%%%%%

In this section, we perform order-of-magnitude estimates of the excess energy accumulated in three components of the climate system (atmosphere, oceans, land).  Due to its importance to humans, we discuss the excess energy in the atmosphere in some details (the other two components are analyzed in a similar way).  We do not consider the warming of the cryosphere (melting of ice) here,  since we do not know of a simple way to relate it to a $1\degree$C increase in global mean temperature.

%%%%%%%%%%%%%%%%%%%%%%%%%%%%%%%%%%%%%%%%%%%%%%%%%%%%%%%%%%%%%%%%%%%%%%%%%%%%%%
\subsection{Global increase in the energy content of the atmosphere}
\label{sec:Global_increase_energy_atmosphere}
%%%%%%%%%%%%%%%%%%%%%%%%%%%%%%%%%%%%%%%%%%%%%%%%%%%%%%%%%%%%%%%%%%%%%%%%%%%%%%

When heating water, one needs to provide energy.  There is of course a difference between heating a pot of water to make tea versus heating a whole swimming pool.  Said differently, the amount of energy required is proportional to the amount of water to heat.  The same can be said of any substance.

\vspace{\baselineskip}

Our interest here is the heating of the atmosphere by the Sun.  When the Sun (directly and indirectly) heats the atmosphere, part of the energy is trapped by it, and another part is radiated back into space via thermal radiation, maintaining the atmosphere out-of-equilibrium.  Due to human emissions of greenhouse gases, an additional amount of energy is trapped in the atmosphere, thus increasing its global mean temperature.  To estimate how much additional energy is trapped, we use:
\begin{eqnarray}
\label{eq:Heat_capacity}
\Delta Q_{\rm atm} & = & m_{\rm air} c_{p} \Delta T,
\end{eqnarray}
where $\Delta Q_{\rm atm}$ represents the heat necessary to increase the temperature of a mass of air $m_{\rm air}$  in the atmosphere by an amount $\Delta T$, and $c_{p}$ is the specific heat at constant pressure for air (we assume the heating is done at constant pressure, since the volume of air in the atmosphere is unconstrained).

\vspace{\baselineskip}

For the purpose of this order-of-magnitude estimate, we need to estimate the mass of air  that is heating up due to greenhouse gas emissions.  This mass of air can be written as:
\begin{eqnarray}
m_{\rm air} & = & \rho_{\rm air} V_{\rm air},
\end{eqnarray}
assuming that the density of air $\rho_{\rm air}$ is constant within the volume of air $V_{\rm air}$ heating up due to greenhouse gases emissions (a crude approximation, but sufficient for our estimate).  According to infrared radiative transfer theory~\cite{Pierrehumbert_2010,Pierrehumbert_2011}, the Earth re-emits a part of the  (visible) incoming solar radiation in the form of infrared radiation at its radiative layer.  The radiative layer is located at an altitude $h_{\rm rad}$ where the concentration of greenhouse gases is sufficiently low for the atmosphere to be optically transparent to infrared radiation, which is about $5$~km [this is the altitude where air is at $255$ K, which is the blackbody temperature the Earth would have in the absence of an atmosphere,  based on the energy-balance equation].  The volume of air heating up due to greenhouse gas emissions is thus:
\begin{eqnarray}
V_{\rm air} & \approx & \frac{4\pi (R_{E} + h_{\rm rad})^{3}}{3} - \frac{4\pi R_{E}^{3}}{3} \;=\; 2.55 \times 10^{18} \;\mbox{m}^{3},
\end{eqnarray}
where $R_{E} = 6371$ km is the radius of the Earth.  The density of air varies with temperature, and air temperature below the radiative layer varies from $288$ K at sea level to $255$ K at the upper boundary.  For this estimate, we take $\rho_{\rm air} = 1.293$ kg/m$^{3}$ (i.e. the density of dry air at $273$ K and $101.3$ kPa)~\cite{Tsonis_2007} as the average density of the atmosphere below the radiative layer.  Using the above values, we obtain $m_{\rm air} \approx 3.30\times 10^{18}$ kg.

\vspace{\baselineskip}

Using the value $c_{p} = 1005$ J/(kg$\cdot$K) for dry air at $273$ K, we can estimate the heat necessary to increase the temperature of the relevant atmospheric volume by $1$~K using Eq.~\ref{eq:Heat_capacity}, giving $\Delta Q_{\rm atm} \approx 2.56\times 10^{21}$ J.  By everyday standards, this is a large amount of energy.    For comparison, the world energy consumption in 2019 was $601$ quad BTU (or $6.34\times 10^{20}$ J) according to the U.S. Energy Information Administration~\cite{EIA}.  Thus the excess energy trapped in the atmosphere due to a one-degree Celsius increase in global mean temperature corresponds to approximately four times the yearly world energy consumption.  

\vspace{\baselineskip}

According to the IPCC reports~\cite{IPCC_2021_PHysical_Science_Basis}, the total increase in global energy content in the period 1971-2018 (from direct temperature measurements) is $4.34\times 10^{23}$ J.  Of this total amount, about $1\%$ goes into heating the atmosphere, corresponding to $4.34\times 10^{21}$~J.  Thus despite being very crude, our order-of-magnitude estimate is within a factor of two of the findings of the IPCC.

%%%%%%%%%%%%%%%%%%%%%%%%%%%%%%%%%%%%%%%%%%%%%%%%%%%%%%%%%%%%%%%%%%%%%%%%%%%%%%
\subsection{Global increase in the energy content of oceans and land}
\label{sec:Global_increase_energy_oceans_land}
%%%%%%%%%%%%%%%%%%%%%%%%%%%%%%%%%%%%%%%%%%%%%%%%%%%%%%%%%%%%%%%%%%%%%%%%%%%%%%

The excess energy accumulated in oceans and land can also be estimated using Eq.~\ref{eq:Heat_capacity}.  First we note that all three climate components (oceans, land, atmosphere) are in thermal contact and constantly exchange heat with each other at their boundaries.  Thermal diffusion (in addition to convection for oceans) allow heat to propagate to greater depths over time.  This is a slow process, and data shows that ocean temperature on the surface has increased more than at greater depths~\cite{vonSchuckmann_etal_2020}.  For the purpose of this estimate, we concentrate on the increase in temperature in the surface layer (i.e. first $20$ m) of oceans and land, and assume that both have warmed up as much as the atmosphere (i.e. $\Delta T \sim 1$ K) \cite{NOAA_oceans_land_temperature,EPA_climate_change_indicator}.  Since our assumption neglects a large volume of oceans and land, we expect our calculations to underestimate the total amount of excess energy in those climate components.  And since heat propagation is more efficient in oceans due to convection, we expect the underestimate to be larger for oceans than for land.

\vspace{\baselineskip}

Based on the above assumption, we estimate the volume of water warming up in the oceans as:
\begin{eqnarray}
\label{eq:Volume_oceans}
V_{\rm water} & \approx & 0.71 \left(4\pi R_{E}^{2}\right) \times h_{\rm surface\;layer} \;=\; 7.24\times 10^{15}\;\mbox{m}^{3},
\end{eqnarray}
where $h_{\rm surface\;layer} = 20$ m and the $0.71$ factor represents the fraction of the Earth's surface covered by oceans.  Taking the density and specific heat of water to be $\rho_{\rm water} = 997$ kg/m$^{3}$ and $c_{p} = 4186$ J/(kg$\cdot$K) respectively, the approximate amount of excess energy accumulated in the oceans is $\Delta Q_{\rm oceans} \approx 3.02\times 10^{22}$ J.  Similarly, we can estimate the volume of surface rocks warming up using Eq.~\ref{eq:Volume_oceans} (with the replacement $0.71\rightarrow 0.29$ to account for the difference in land and ocean surface area), giving $V_{\rm rock} \approx 2.96\times 10^{15}$~m$^{3}$.  Taking the average density and specific heat of rocks in the crust to be $\rho_{\rm rock} \approx 2550$ kg/m$^{3}$ \cite{USGS} and $c_{p}\approx 800$ J/(kg$\cdot$K) \cite{Waples_2004}, we obtain $\Delta Q_{\rm land} \approx 6.04\times 10^{21}$ J for the amount of excess energy accumulated in land.  

\vspace{\baselineskip}

From the above order-of-magnitude estimates, we already note that the amount of excess energy accumulated in oceans (and to a lesser extent land) is much larger than the one in the atmosphere ($\Delta Q_{\rm oceans}\sim 12\Delta Q_{\rm atm}$ and $\Delta Q_{\rm land}\sim 2.4\Delta Q_{\rm atm}$).  This large difference in energy content between oceans and atmosphere or land is mainly due to the large specific heat of water compared to dry air or rock, and the total mass of each component involved.  A comparison between our estimates and the more detailed analysis presented in the IPCC reports~\cite{IPCC_2021_PHysical_Science_Basis} show that our simple calculations capture the correct ordering for the excess energy in each component (i.e. $\Delta Q_{\rm oceans}\sim 5\Delta Q_{\rm land} \sim 12\Delta Q_{\rm atm}$ versus $\Delta Q_{\rm oceans}^{\rm (IPCC)}\sim 18\Delta Q_{\rm land}^{\rm (IPCC)} \sim 90\Delta Q_{\rm atm}^{\rm (IPCC)}$), but underestimates the energy contained in the oceans by more than an order of magnitude (i.e. $\Delta Q_{\rm oceans}^{\rm (IPCC)} \sim 13\Delta Q_{\rm oceans}$).  This underestimation is expected, given our initial assumption of neglecting the warming of water at depths greater than $20$ m.

%%%%%%%%%%%%%%%%%%%%%%%%%%%%%%%%%%%%%%%%%%%%%%%%%%%%%%%%%%%%%%%%%%%%%%%%%%%%%%
\section{Excess global energy content and hurricanes}
\label{eq:Global_energy_hurricanes}
%%%%%%%%%%%%%%%%%%%%%%%%%%%%%%%%%%%%%%%%%%%%%%%%%%%%%%%%%%%%%%%%%%%%%%%%%%%%%%

To establish a cause-and-effect relationship between a larger (human induced) global mean temperature and extreme meteorological events is a difficult task that requires an understanding of the physics of the climate system and its natural variability over many spatial and temporal scales~\cite{Trenberth_etal_2015}.  This is beyond the scope of the present paper, and we thus need to focus on the more modest goal of providing plausibility arguments that a larger global mean temperature could lead to more extreme meteorological events.  %In order to do so, we estimate the energy required to produce and maintain a typical hurricane over its lifetime.

\vspace{\baselineskip}

Energy is required to power (extreme) meteorological events, such as hurricanes.  Thus our task is to show quantitatively that the climate system has enough excess energy in its budget to power additional hurricanes. Said differently, we estimate the energy necessary to form and maintain a hurricane, and compare this to the excess energy found in Sect.~\ref{sec:Global_increase_energy}.  Note that hurricanes vary enormously in size and strength.  For the purpose of this estimate, we use ``typical'' or ``average'' hurricane-related quantities coming from a variety of sources~\cite{NOAA_hurricane_FAQs,Unidata,Hurricane_Science}.  We approximate a hurricane as a cylinder of circulating air (diameter of $500$ km) with an ``eye'' (diameter of $50$ km) and a height of $15$ km.  We assume the wind speed is zero in the eye, and constant in the rest of the cylinder.  For a hurricane to form, the air in the cylinder must go from rest to an average wind speed of $150$ km/h (for a category 2 hurricane).  The kinetic energy necessary to form an ``average'' hurricane is thus:
\begin{eqnarray}
E_{k} & = & \frac{1}{2}m_{\rm hurricane} v^{2} \;=\; \frac{1}{2}\left(\rho_{\rm air} V_{\rm hurricane}\right) v^{2} \;\approx\; 3.28 \times 10^{18} \;\mbox{J},
\end{eqnarray}
where $m_{\rm hurricane}$ is the mass of air in the hurricane, $V_{\rm hurricane}$ is the volume of the hurricane, $v$ is the average wind speed, and we used the same air density as in Sect.~\ref{sec:Global_increase_energy_atmosphere}.  A hurricane forms from a tropical depression in a few days.  Taking an average formation time of $5$ days, the power required to form a hurricane is $3.28 \times 10^{18}\;\mbox{J}/5\;\mbox{days} \approx 7.59\times 10^{12}$~W.  Our estimate is consistent with estimates on the power required to maintain a hurricane~\cite{Emanuel_1998}.  A hurricane maintains itself when on water, and can last between one day and one month.  Assuming an average lifetime of two weeks (including its formation time of a few days), the total energy required to form and maintain a hurricane is of the order of $9.18\times 10^{18}$~J.  

\vspace{\baselineskip}

Since hurricanes are manifestations of kinetic energy in the atmosphere, it is tempting to compare the energy necessary to form and maintain a hurricane to the excess energy trapped in the atmosphere due to global warming, and argue that this excess energy goes into the creation of additional hurricanes (note that this approach disregards the physics of hurricanes and should be taken with caution, see Sect.~\ref{sec:Discussion} for a discussion).  Doing this comparison, we see that the excess energy in the atmosphere is sufficient to power roughly $278$ hurricanes.  

\vspace{\baselineskip}

Even if the estimates presented here are crude, they correlate with data on hurricanes and global warming.  Statistics on the number of hurricanes~\cite{NOAA_hurricane_statistics} seem to indicate that the average number of hurricanes per decade steadily increases: $66$ in the 70's, $69$ in the 80's, $89$ in the 90's, $110$ in the 2000's, and $100$ in the 2010's.  Taking the 70's as a baseline (when satellite imagery allowed an accurate hurricane count and global temperature started to increase), we see that there is an ``excess'' of $104$ hurricanes compared to the baseline.  Since global mean temperature increased by $1\degree$ C since the 70's (corresponding to the energy of roughly $278$ hurricanes), we conclude that the  excess trapped energy in the atmosphere could in principle be sufficient to power the ``excess'' hurricanes in the Atlantic basin since the 70's.

%%%%%%%%%%%%%%%%%%%%%%%%%%%%%%%%%%%%%%%%%%%%%%%%%%%%%%%%%%%%%%%%%%%%%%%%%%%%%%
\section{Discussion}
\label{sec:Discussion}
%%%%%%%%%%%%%%%%%%%%%%%%%%%%%%%%%%%%%%%%%%%%%%%%%%%%%%%%%%%%%%%%%%%%%%%%%%%%%%

The simple numerical estimates for the excess energy trapped in the climate system and energy required to power hurricanes presented in Sect.~\ref{sec:Global_increase_energy}-\ref{eq:Global_energy_hurricanes} are consistent with other estimates found in the literature, but it is important to emphasize that the correlation between the increase in global temperature and the increase in the number of hurricanes should be taken with caution.  Here are some comments on this last point:

\begin{itemize}
	\item According to the IPCC reports~\cite{IPCC_2021_PHysical_Science_Basis}, ``there is low confidence in long-term (multi-decadal to centennial) trends in the frequency of all-category tropical cyclones'', and ``it is likely that the global proportion of major (Category 3–5) tropical cyclone occurrence has increased over the last four decades.''  This indicates that the increase in frequency and intensity of hurricanes and tropical cyclones are not considered to be the most robust indicators of human induced global warming.  Two important factors contributing to this lack of robustness are the difficulty of accounting for the natural variability of the climate system over decades and centuries (i.e. difficulty of separating natural trends from human induced ones), and the lack of reliable data on tropical cyclones prior to satellite imagery.  This implies that the correlation presented at the end of Sect.~\ref{eq:Global_energy_hurricanes} might simply be a fluke due to the natural variability of the climate system, although a recent detailed statistical analysis of climate model simulations~\cite{Murakami_etal_2020} shows that there is a global spatial trend in the frequency of tropical cyclones.  This observed trend (i.e. decrease in the southern Indian Ocean and western North Pacific and increase in the North Atlantic and central Pacific) is very unlikely to be explained using natural multi-decadal variability, and may be partly attributable to the increase in greenhouse gases.  

	\item Hurricanes are powered by the release of latent heat as evaporated water from the ocean condenses in the atmosphere~\cite{Smith_2006}.  They can be viewed as Carnot engines, with an efficiency related to the difference between the sea-surface temperature and the mean temperature at which heat is exported by or lost from the hurricane's high-level outflow~\cite{Emanuel_1991}.  Even if atmospheric temperature may play a role in the physics of hurricanes, it is not the main driver behind their formation and maintenance.  Thus the correlation presented at the end of Sect.~\ref{eq:Global_energy_hurricanes} might be fortuitous.  Even if an increase in mean global atmospheric temperature is connected to an increase in mean sea-surface temperature, it is the energy content of oceans (and not the atmosphere) that plays a major role in hurricane physics~\cite{Trenberth_etal_2018}.  But as demonstrated in Sect.~\ref{sec:Global_increase_energy_oceans_land} using simple estimates and experimentally quantified using in situ measurements and satellite imagery~\cite{IPCC_2021_PHysical_Science_Basis,vonSchuckmann_etal_2020}, oceans contribute $91\%$ to the total heating of the whole climate system (versus only $1\%$ for the atmosphere), showing that the excess energy trapped in the climate system is more than enough to power thousands of hurricanes and other extreme weather events.  

	\item By itself, a higher thermal energy content in the climate system does not necessarily imply more extreme meteorological events (i.e. larger ``fluctuations'').  This excess thermal energy must be transformed into useful work through some physical mechanism.  Any plausibility argument based on the estimates presented in this paper must be supplemented by arguments connecting a higher thermal content in the climate system to the occurrence of larger fluctuations in the system.  To understand this, one can use the following simple analogy.  Imagine a circular pool with a water depth of $1$ m, and little rubber ducks floating on the water.  In the absence of external forcing, the water surface stays flat (in equilibrium), and no work is done on the rubber ducks.  If external forcing is applied (e.g. moving the pool walls), waves of all sizes are produced and do work on the rubber ducks.  An extreme event (large fluctuation) in which most of the water in the pool is concentrated in the center is possible, but less likely than smaller waves.  The maximum size of this extreme wave is limited by the total amount of water in the pool.  Now increasing the total amount of water in the pool (say by $1$ cm) has the potential of increasing the size of the maximum wave, as well as the likelihood of producing waves smaller than this new maximum wave.  Note that the presence of external forcing is necessary to get these (extreme) waves, and it is the waves that are doing the work on the rubber ducks.  Similarly, an atmosphere at a uniform temperature does not in itself produce extreme meteorological events, even if the temperature is very high.  External forcing (e.g. from the Sun) is necessary to produce those extreme events.  But a higher temperature for the atmosphere has the potential of producing larger fluctuations (just like having more water in the pool has the potential of producing larger waves).

\end{itemize}

%%%%%%%%%%%%%%%%%%%%%%%%%%%%%%%%%%%%%%%%%%%%%%%%%%%%%%%%%%%%%%%%%%%%%%%%%%%%%%
\section{Conclusion}
\label{sec:Conclusion}
%%%%%%%%%%%%%%%%%%%%%%%%%%%%%%%%%%%%%%%%%%%%%%%%%%%%%%%%%%%%%%%%%%%%%%%%%%%%%%

In this paper, we estimate that the excess thermal energy trapped in the climate system (oceans, land, atmosphere) due to human induced global warming since the 70's is approximately $3.88\times 10^{22}$~J, with the largest part absorbed by the oceans ($3.02\times 10^{22}$~J) and the smallest part taken up by the atmosphere ($2.56\times 10^{21}$ J).  Even if our order-of-magnitude estimates are based on simple physics and a single ``climate related'' number (i.e. a $1\degree$C increase in global temperature), they are in good agreement with the findings of the IPCC.

\vspace{\baselineskip}

This huge excess amount of energy in the three climate components is {\em in principle} sufficient to power multiple hurricanes (estimated to require about $9.18\times 10^{18}$ J of energy over their lifetime), although establishing a cause-and-effect relationship between the two requires an understanding of the complex physics of the climate system (which is beyond the scope of this paper).  From a purely energetic point of view however, our estimates show that it is at least plausible that a large number of extreme weather events (such as hurricanes) could be powered by global warming.  The arguments presented here are sufficiently simple to be presented in introductory physics classes, and we hope they could help bridge a conceptual gap in the public mind between a seemingly small increase in mean global temperature and the increase in extreme weather events.

\ack

The author would like to thank D. Blowers from the Castleton Spartan for asking him the question that started him thinking about the topic presented in this paper, and Mark R. Petersen for useful comments and suggestions.

\vspace{\baselineskip}

\bibliography{bibliography_file} % Use the bibliography.bib file for the bibliography
\bibliographystyle{unsrt} % Use the plainnat style of referencing

\end{document}